\documentclass[prb,amsmath,amssymb,twocolumn]{revtex4}
\usepackage[normalem]{ulem}
\usepackage{amsmath,amssymb}
\usepackage[bookmarks=true,colorlinks,linkcolor=blue,urlcolor=blue,citecolor=blue]{hyperref}
\usepackage{graphicx,amsmath,amsfonts}
\usepackage[usenames]{color}
\usepackage{epstopdf}
\usepackage{verbatim}
\usepackage{amsthm}

\newcommand{\be}{\begin{equation}}
\newcommand{\beq}{\begin{eqnarray}}
\newcommand{\eeq}{\end{eqnarray}}

\def \be{\begin{equation}}
\def \ee{\end{equation}}
\def \ba{\begin{array}}
\def \ea{\end{array}}
\def \bea{\begin{eqnarray}}
\def \eea{\end{eqnarray}}
\def \nn{\nonumber}
\def \half{\frac{1}{2}}

\def \W{{\Omega}}
\def \e{{\epsilon}}

\def \a{{\alpha}}

\def \d{{\delta}}
\def \w{{\omega}}
\def \s{{\sigma}}
\def \f{{\varphi}}

\def \e{{\epsilon}}

\def \nd{{^{\vphantom{\dagger}}}}
\def \yd{^\dagger}
\def \av#1{{\langle#1\rangle}}

\begin{document}

\title{The superfluid insulator transition of ultra-cold bosons in disordered 1d traps}
\author{Ronen Vosk and Ehud Altman}
\affiliation{Department of Condensed Matter Physics, The
Weizmann Institute of Science, Rehovot, 76100, Israel}
\date{\today}

\begin{abstract}
We derive an effective quantum Josephson array model for a weakly interacting one-dimensional condensate that is fragmented into weakly coupled puddles by a disorder potential. The distribution of coupling constants, obtained from first principles, indicate that weakly interacting bosons in a disorder potential undergo a superfluid insulator transition controlled by a strong randomness fixed point [Phys. Rev. Lett. 93, 150402 (2004)]. We compute renormalization group flows for concrete realizations of the disorder potential to facilitate finite size scaling of experimental results and allow comparison to the behavior dictated by the strong randomness fixed point. The phase diagram of the system is obtained with corrections to mean-field results.
\end{abstract}

\maketitle
\section{Introduction}
A number of recent experiments have investigated the properties of Bose condensates in disordered traps\cite{Billy2008,Inguscio,Chen2008a} and thereby revived the theoretical interest\cite{AKPR,Lugan2007,Falco2009,Aleiner2010} in a fundamental unsolved problem:
what is the fate of Anderson localization in the presence of interactions and strong quantum correlations?
The problem is particularly intriguing in one dimension where disorder alone or interactions alone would have a profound effect on the physics. %Interactions immediately destroy the long range phase coherence of a zero temperature condensate, while any amount of disorder causes localization of the single particle eigenstates.

One way to tackle the problem is to start from the harmonic fluid (Luttinger liquid) description of a uniform interacting Bose gas\cite{Haldane1981}, then add disorder to it as a small perturbation. Using this approach Giamarchi and Schultz\cite{Giamarchi1988} predicted a superfluid-insulator quantum phase transition that occurs at a universal value of the Luttinger parameter, or correlation decay exponent. This approach is justified a priori if the chemical potential set by the interactions is much larger than the disorder strength. However, most experiments are in the opposite limit of weak interactions, where the disorder acts to fragment the condensate invalidating the Luttinger liquid description. The non-interacting ground state is also a bad starting point. This state, in which all particles occupy the lowest single-particle localized state,
is unstable to adding even the weakest interaction. The absence of a simple basis from which to formulate a perturbation expansion makes the limit of weakly interacting bosons inherently strongly correlated.

\begin{figure}
 \centerline{\resizebox{0.9\linewidth}{!}{\includegraphics{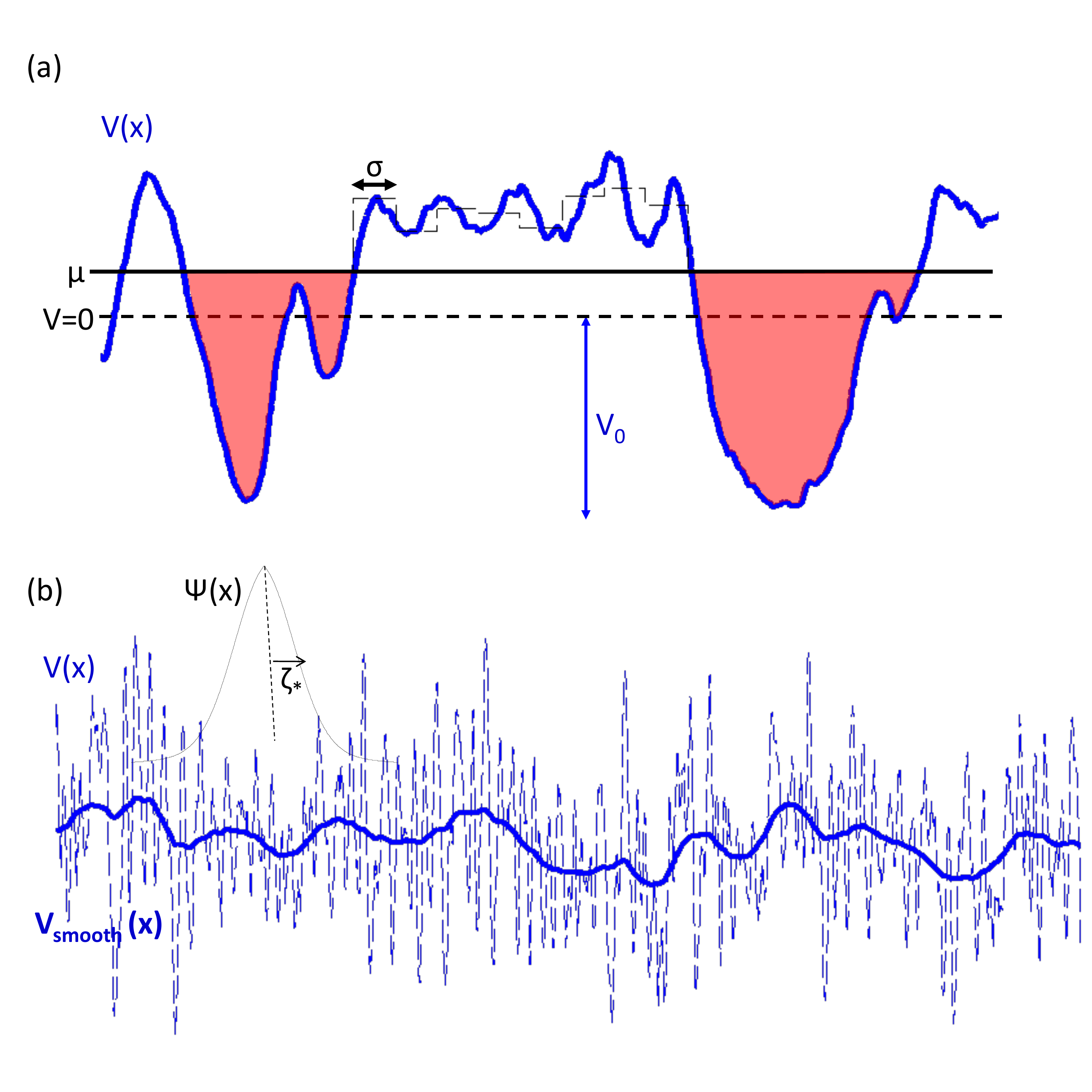}}}
 \caption{{\em Schematic sketch of the disorder potential and condensate fragmentation.} (a) Regime of smooth disorder -- typical puddles are much larger than the healing length. (b) A rough disorder is effectively smoothed by the healing length or single particle localization length. In both cases the distribution of weak links is determined by atypical long barriers.}
 \label{fig:setup}
\end{figure}

In this paper we derive from first principles a low energy effective model of the disordered quantum gases in the form of a random Josephson junction array. Such a model was assumed in previous work, coauthored by one of us\cite{AKPR,AKPR2,AKPR3}, as a starting point for a real-space renormalization group (RSRG) analysis\cite{Dasgupta1980,Fisher1994}. This analysis predicts a superfluid insulator transition controlled by a strong randomness fixed point, distinct from the transition described by Giamarchi and Schulz\cite{Giamarchi1988}. However the connection between the quantum gas in the disorder potential and the effective Josephson array model has not been established.

Starting from the microscopic random potential for the atoms, we show how the condensate fragments into mezoscopic puddles. We derive the distribution of Josephson couplings between the puddles and of the charging energies within them. This is done for two different regimes of the disorder potential: rough and smooth potential as compared to the healing length and to the single particle localization length of the condensate (see Fig. \ref{fig:setup}). Remarkably in both cases the distributions calculated from first principles, are precisely in the form of the stable solutions of the RG equations found in Ref. \onlinecite{AKPR}. In particular the distribution of Josephson links $P(J)$ behaves as a power law of $J$ at small $J$. Hence given a set of microscopic parameters we can immediately compute the phase diagram and make direct predictions for finite size scaling of observables. % The analysis confirms that weakly interacting bosons in a disordered potential undergo a superfluid insulating transition controlled by the strong randomness fixed point.

\section{The Model and Mapping to Josephson Array model}
Our starting point for the theoretical analysis is the continuum boson hamiltonian
\be
H=\int dx\psi\yd\left(-{\nabla^2\over 2m}-V(x)-\mu\right)\psi\nd+u\psi\yd\psi\yd\psi\nd\psi\nd
\label{Hmic}
\ee
Here $u$ is the effective contact interaction and $V(x)$ is a random potential assumed to be gaussian and characterized by the auto-correlation function $\av{V(x)V(x')}\approx V_0^2 e^{-|x-x'|/\s}$.

For a given chemical potential, the potential landscape is filled with particles up to the chemical potential, forming local superfluid puddles (see Fig. \ref{fig:setup}(a)). Each puddle is assumed to be characterized by a single phase $\varphi_i$ and occupation number $N_i$ which are non-commutating operators. We would like to map the physics of this system to an effective random Josephson junction array model
\be
\label{eq:JJA_hamiltonian}
    H=\frac{1}{2} \sum_i U_i N_i^2 - \sum_i J_i cos\left(\varphi_{i+1} - \varphi_{i}\right)
\ee
Here $U_i$ is the inverse capacitance, or charging energy of a of puddle $i$ and $J_i$ is the Josephson coupling between puddles $i$ and $i+1$.

  The Josephson couplings can be obtained from the mean field Gross-Pitaevskii (GP) solution of (\ref{Hmic}). The formula for the superfluid stiffness of a one-dimensional condensate in terms of the GP wavefunction $\psi$ is (see appendix \ref{sec:stiffness} and Ref. \onlinecite{Fontanesi2011})
\be
\rho_{MF}^{-1}={1\over L}{m\over \hbar^2}\int_0^L dx {1\over |\psi|^2 }
\label{rhos}
\ee
This integral is dominated by deep minima of $\psi$ that occur between neighboring puddles.
We can compute the stiffness as a sum over those saddle point contributions
$\rho_{MF}^{-1}=1/N_{sp}\sum J_i^{-1}$, where the $J_i$ are interpreted as the Josephson
couplings between puddles and $N_{sp}$ is the number of deep minima. More generally, the integral (\ref{rhos}) can be subdivided
into sub regions of order of the correlation of $|\psi(x)|^2$
to obtain a distribution of local phase stiffness. The same analysis can be used to obtain the effective Josephson array from a density profile measured in experiment. We emphasize that this is a method to obtain the local phase stiffness $J_i$ between puddles, which enter the effective Hamiltonian (\ref{eq:JJA_hamiltonian}). The true thermodynamic stiffness can then be calculated only within the quantum Hamiltonian (\ref{eq:JJA_hamiltonian}). In general it will be renormalized downward compared to the mean field  stiffness (\ref{rhos}) because of phase slips induced by the charging terms $U_i$.

We shall compute $\psi(x)$  using a simple approximation to the GP equation, which allows us to obtain analytic results for the distributions of coupling constants. The results will be checked against the distributions obtained by exact numerical calculation of the GP ground state.

We distinguish two regimes according to the ratio between the length scale $\s$ of disorder potential fluctuations to the
natural correlation length on which the condensate amplitude $|\psi|^2$ can adopt to the external changes.
The latter is determined by the minimum of two natural scales: (i) the localization length of
non interacting particles at zero energy $\zeta_*= (\hbar^4/V_0^2\s m^2)^{1/3}$ ~\cite{Lifshitz1965,Halperin1966} and (ii)
The healing length of the condensate  $\xi_h=\hbar/\sqrt{m (\mu+V_0)}$.
The case of smooth disorder, where
$\s$ is the largest scale is conceptually somewhat simpler, and we shall therefore start the analysis from this regime. Later we will show that the rough disorder limit can be treated in an analogous way.

\subsection{Smooth disorder} In a smooth potential, the condensate has appreciable amplitude only where the potential dips below the chemical potential. These regions define the superfluid puddles as illustrated in Fig. \ref{fig:setup}(a). Josephson coupling between puddles is induced by tunneling under the potential barriers separating them.

Using Eq. (\ref{rhos}) and the WKB approximation (see appendix \ref{sec:J}) we obtain the coupling constants:
\be
J=\sqrt{{\left|V^\prime_1V^\prime_2\right|\over \pi}}~\frac{0.64\hbar^2}{m u}e^{-{1\over\hbar}\int_{x_{1}}^{x_{2}}dx\sqrt{2m(V(x)-\mu)}}.
\label{eq:JWKB}
\ee
Here $x_i$ denote the two edges of the barrier, defined by $V(x_i)-\mu=0$ and $V^\prime_i\equiv\left(dV/dx\right)_{x=x_i}$. Hence the Josephson coupling is composed as a product of three random variables $J\sim y_1 y_2 T$, where $y_i=\sqrt{\left|V^\prime_i\right|}$ and $T$ is the exponential factor.

The macroscopic stiffness of the chain is determined by weak Josephson links that arise from atypically large barriers, much longer than the disorder correlation length $\s$. For such barriers, the variables $y_1\text{,}$ which depends on the  left edge of the barrier,$y_2$ which depends on the right edge of it, and  $T$ which depends on the potential in the bulk of the barrier, are essentially independent. We can therefore obtain the distribution of each of the three variables separately in order to construct the distribution of $J$.

To compute the statistics of the exponential factor in (\ref{eq:JWKB}) for long barriers, we can split the integral to a sum on segments of size $\s$ on which the potential is approximately constant
\be
I=\int_{x_1}^{x_2} dx\sqrt{2m(V(x)-\mu)}\approx\s\sum_{i=1}^l\sqrt{2m(V_i-\mu)}.
\label{Isum}
\ee
Here $V_i>\mu$ are independent random variables, distributed as
\be
P_\mu(V)={1\over q_\mu \sqrt{2\pi V_0^2}}e^{-V^2/2V_0^2}.
\ee
$q_\mu=(2\pi)^{-1/2}\int_{\mu/V_0}^{\infty} dy e^{-y^2/2}$ is the probability to find a potential $V(x)>\mu$.
At the same time, the probability for a barrier of length $l$ is $p(l)=q_\mu^{l-1} (1-q_\mu)\approx |\ln q_\mu| \exp(-|\ln q_\mu|l)$, that is, the probability of having $l$ consecutive segments with $V_i>\mu$. Note that in the $2^{\text{nd}}$ (approximate) equality we moved to a continuous $l$ while keeping the distribution normalized.

  The tail of the distribution of the sum $I$ is dominated by large $l$ (not by large $V$ since the distribution of $V$ has a much faster decay). We can therefore apply the central limit theorem
%The probability distribution of barriers $I$ is simple to write in terms of the characteristic functions $\f(t)$ of the $V$ distribution
%\be
%P(I)=\int_{-\infty}^\infty dt \sum_l \f(t)^l p(l) e^{-it I},
%\ee
to express $I$ as a function of just two independent random variables $I(l,\eta)=\s (\kappa l+ \eta\d\kappa \sqrt{l})$, where $\kappa$ and $\d\kappa$ are the average and
standard deviation respectively of $\sqrt{ 2m(V-\mu)}$ over the distribution $P_\mu(V)$. $\eta$ is a gaussian variable with zero mean and unit variance.

Given the distributions of $l$ and $\eta$ it is a straight forward exercise to compute the probability distribution of the exponential factor $T=\exp\left(-I(\eta,l)/\hbar\right)$. In the limit of small $T$ we find $P(T)= A T^\chi$, with the exponent $\chi$
\be
\chi=-1+\frac{1}{\sqrt{2}}\left({\zeta_*\over \s}\right)^{3\over 4} \d\tilde{\kappa}^{-2}\left(\sqrt{{\tilde{\kappa}}^2-2\d\tilde{\kappa}^2\ln q_\mu }-\tilde{\kappa}\right)
\label{alpha-result}
\ee
and the pre-factor
\be
A = \frac{-\ln q_\mu}{\sqrt{2}} \left(\frac{\zeta_*}{\sigma}\right)^{3/4} \left({\tilde{\kappa}}^2-2\d\tilde{\kappa}^2\ln q_\mu\right)^{-1/2},
\ee
as expressed in terms of the rescaled parameters $\kappa=\sqrt{2m V_0}\tilde{\kappa}(\mu/V_0)$ and $\d\kappa=\sqrt{2mV_0}\d\tilde{\kappa}(\mu/V_0)$.
A rigorous derivation of the distribution $P(T)$ (see appendix \ref{sec:PJ}) gives essentially the same result as the more heuristic derivation presented above.

The distribution of Josephson couplings may depend also on the distribution of the pre-factors $y_i=\sqrt{\left|V^\prime_i\right|}$. Because the potential is a gaussian variable, $y_i$ is distributed as $p(y)\propto y$ at small values of $y$.
Consequently, the distribution of Josephson coupling at small values of $J$ is given by $J=y_1 y_2 T=(A/\W_0)(J/\W_0)^\a$
  with $\a=min(\chi,1)$ . The power-law holds up to the cutoff scale $\W_0$ determined by the pre-factor of (\ref{eq:JWKB}). Using the typical value $V^\prime\sim V_0/\s$  we have $\W_0\approx 0.4(\hbar^2/m)(V_0/ \s u)$.

Note that some of the junctions in the effective model are actually formed with $J>\W_0$ and therefore lie beyond the power-law distribution. However puddles separated by these strong junctions can be joined to make larger effective puddles in the same spirit of the real-space $RG$\cite{AKPR}. If the charging energies are concentrated well below $\Omega_0$, which we shall see is the natural situation in experiments, then the process of removing strong links does not modify the exponent $\a$ of the distribution below $\Omega_0$ and we only obtain a normalized distribution for $J\in[0,\W_0]$:
\be
P_0(J)= \W_0^{-1}(\a+1)(J/\W_0)^{\a}.
\label{PJ}
\ee

When comparing to experiments it is important to take into consideration the physical length of the condensate and translate it to the number of junctions in the effective Josephson array. By dividing the total length $L$ to the average size of a junction and barrier we find the number $N_*=(L/\sigma)q_\mu(1-q_\mu)$. But as noted, some of the junctions lie above the cutoff $\Omega_0$. After removing the strong junctions, as described above, we are left with $N_0=N_*\int_0^{\Omega_0} dJ P(J)=N_* A\left(\alpha+1\right)$ junctions which follow the pure power-law distribution (\ref{PJ}).
This number turns out to be only slightly ($\lesssim 10 $) lower than the total number of junctions $N_*$ for chemical potentials $\mu\lesssim V_0$. In paractice it is therefore sufficient to take $N_*$ as the starting number of junctions.

The charging energy of a puddle can be found using Thomas-Fermi approximation for the wavefunction of the puddle $\psi_i(x)=\sqrt{\left[\mu-V(x)\right]/u}$. The total number of particles is $N_i = \int_{x_{1,i}}^{x_{2,i}}|\psi_i|^2 dx$ and the total energy of the puddle is $E_i = \frac{u}{2}\int_{x_{1,i}}^{x_{2,i}}|\psi_i|^4 dx$ where $x_{1,i}$ and $x_{2,i}$ are the edges of the puddle.
The charging energy is related to the chemical potential by $\mu= U_i N_i$. Using the equations for $N_i$ and $E_i$ once can show that
\be
\label{eq:charging_energy}
\begin{split}
U_i &= N_i^{-1}  \frac{\partial E_i}{\partial{N_i}} =  N_i^{-1} \frac{\partial E_i}{\partial{\mu}} \left[\frac{\partial N_i}{\partial{\mu}}\right]^{-1} \\
& = u/L_i
\end{split}
\ee
which depends on the length of the puddle $L_i$ but not on the puddle's shape.

We can compute the distribution of charging energies $U_i$ in the array from that of the lengths $L_i$ of the superfluid puddles. Since puddles are exactly complementary to barriers, the distribution of puddle sizes is the same as that computed above for the barrier lengths with $q_\mu$ replaced by $1-q_\mu$. The resulting distribution of charging energies is
\bea
F_0(U)&=&{f_0\over \W_0}\left({\W_0\over U}\right)^2e^{-f_0 \W_0/U +f_0}\nn\\
f_0&=&-(u/\s \W_0)\ln(1-q_\mu)
\label{U-distribution}
\eea
where the cutoff scale $\W_0$ was estimated above.

The distributions of the charging energies (\ref{U-distribution}) and of the of the Josephson coupling constants (\ref{PJ}) are precisely the stable solutions of the RG equations for the disorder distributions \cite{AKPR}. The flow of these distributions, substituted as initial conditions to the RG equations will therefore be greatly simplified: from the full functional flow to a flow of only two parameters $\a$ and $f_0$. Such analysis will be described in section \ref{sec:RGflow}.

Finally we note that the results above are written in terms of the chemical potential, whereas the average density is often easier to obtain from experiments. Given the gaussian distribution of the potential and using the Thomas Fermi approximation it is straight forward to obtain the relation between the two as
\be
\rho = \frac{V_0}{u} \left[\frac{\mu}{V_0} (1-q_\mu) + \frac{1}{\sqrt{2\pi}} e^{-\frac{1}{2} \left( \frac{\mu}{V_0}\right)^2} \right].
\label{EOS}
\ee

\begin{figure}[t]
 \centerline{\resizebox{0.9\linewidth}{!}{\includegraphics{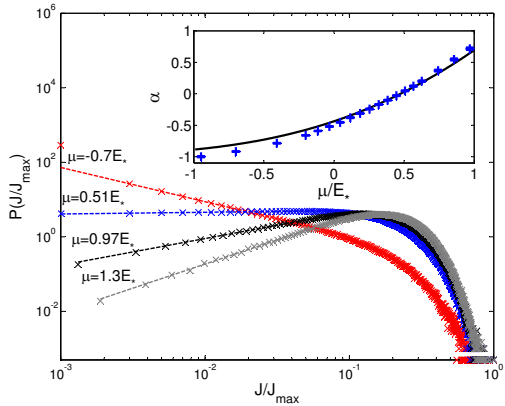}}}
 \caption{Distributions of coupling Josephson couplings computed from exact numerical solution of the Gross-Pitaevskii equation for a condensate in a realistic disorder potential. Inset: the exponent of the power law fit of the low energy part of the distributions is compared to the analytic result using the WKB approximation.}
 \label{fig:exponent}
\end{figure}

\subsection{Rough disorder}
We turn to the case of a rough disorder potential, where $\s\ll \zeta_*,\xi_h$.
Since the potential changes in space faster than the wave-function can respond, it is equivalent to a white noise potential, with the only important parameter being the combination $D=V_0^2\s$.
The disorder strength $D$ is directly related to the localization length at zero energy\cite{Lifshitz1965,Halperin1966} $\zeta_*=(\hbar^4/D m^2)^{1/3}$ and to the characteristic scale $E_*=\hbar^2/2m\zeta_*^2$. That is $D=E_*^2 \zeta_*$.

Since only the combination $V_0^2\s$ matters, while the independent values of $\s$ and $V_0$ are not important, the potential may be replaced by a smoothed effective potential with $\s^{\text{eff}}=\min(\zeta_*,\xi_h)$ and $V_0^{\text{eff}}=V_0\sqrt{\s/\s^{\text{eff}}}$ ~\cite{Sanchez-Palencia2006}. As before we imagine filling this potential landscape, forming weakly coupled superfluid puddles which grow with increasing chemical potential.
In the puddles $\xi_h$ sets the smaller scale when $\mu>0$ whereas $\zeta_*$ is smaller for $\mu<0$. Under barriers on the other hand, $\zeta_*$ is always the relevant length scale. Since we are interested in the vicinity of $\mu=0$ where $\xi_h$ and $\xi_*$ are comparable we can apply
the same analysis as outlined above for the smooth potential, taking $\s\to\zeta_*$ everywhere.  This results in the distributions (\ref{PJ}) and (\ref{U-distribution}) for the coupling constants. At high densities the exponent is $\alpha=1$ and it decreases together with the chemical potential and reaches the critical value $\a=0$ when $\mu\approx E_*/2\equiv \hbar^2/4m\zeta_*^2$.

In order to confirm the approximate analytic result we have solved the GP equation numerically for the case of a rough disorder potential. We find the ground state wave-function using imaginary time propagation and use it in Eq. (\ref{rhos}) to obtain the distribution of Josephson coupling constants shown in Fig. \ref{fig:exponent}(a) for a range of chemical potentials.
The power-law behavior of the distribution at small values of $J$ is seen clearly in the figure. Note that the same analysis, using Eq. (\ref{rhos}), can be done to translate in-situ density profiles measured in experiments to a distribution of Josephson coupling constants.
The inset of Fig. \ref{fig:exponent} shows the exponent of the power-law as a function of the chemical potential derived from both the analytic and the numerical solutions. The approximate analytic result is seen to be in almost perfect agreement with the exact numerical calculation, though the former is rigorously controlled only for positive values of the chemical potential.

\section{RG flow and relation to experiments}\label{sec:RGflow}
We have now derived an effective Josephson array model for bosons in a disorder potential that can be fed into the real-space RG framework of Ref. \onlinecite{AKPR}. The real-space RG consists of gradually eliminating sites with the largest charging energy $U_i$ or Josephson coupling $J_i$. The remaining sites are described by the same Hamiltonian (\ref{eq:JJA_hamiltonian}) with renormalized probability distribution of the charging energies and the Josephson coupling constants.

The flow of the effective disorder with decreasing energy scale is described by a set of integro differential equations for the distributions of coupling constants.
%It was found in Ref. \onlinecite{AKPR} that a specific type of distributions is kept self-similar during the RG flow.
Remarkably the distributions of coupling constants $P_0(J)$ and $F_0(U)$, derived above from the microscopics, are precisely self similar solutions to these equations \cite{AKPR}. Therefore, when feeding these distributions as initial conditions to the RG equations the physics is fully determined by the flow of the two parameters $f_0(\Gamma)$ and $\a(\Gamma)$ with the running RG scale
\be
\begin{split}
\frac{df_0(\Gamma)}{d\Gamma} & = f_0(\Gamma)-f_0(\Gamma)\left[\a(\Gamma)+1\right]\\
\frac{d\a(\Gamma)}{d\Gamma} & = -\left[\a(\Gamma)+1\right] f_0(\Gamma)
\end{split}
\ee
where $\Gamma=\text{log}\left(\Omega_0/\Omega\right)$.

\begin{figure}
 \centerline{\resizebox{0.9\linewidth}{!}{\includegraphics{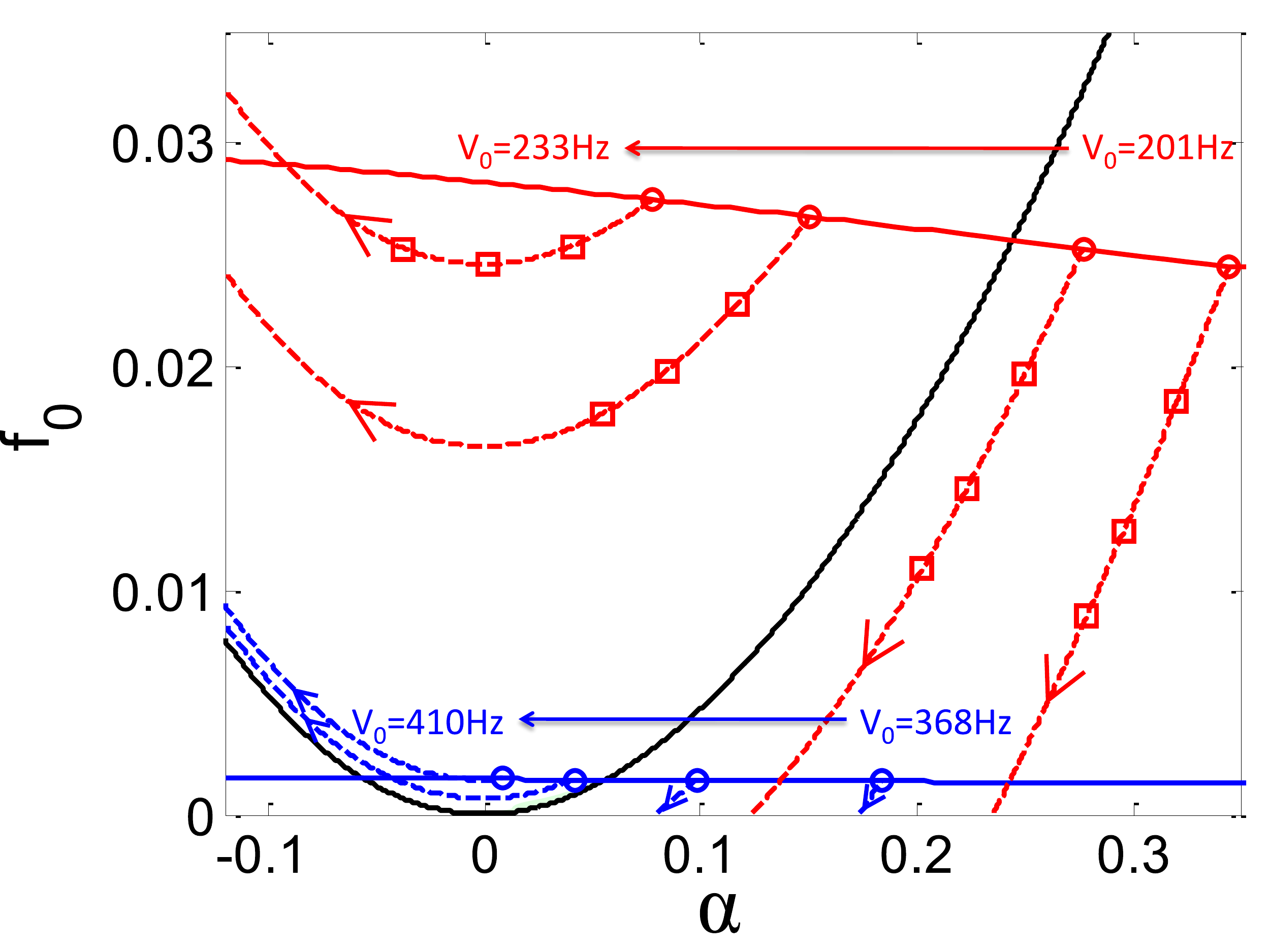}}}
 \caption{{\em Real-space RG flow} obtained from realistic setups with disorder produced by speckle potentials. The two sets of trajectories correspond to different values of the transverse frequency $\w_\perp$ of the quasi one-dimensional trap potential and 1d boson density $\rho$. $\w_\perp=0.66~Khz (2~Khz)$ and $\rho=25 (4)$ atoms per $\mu m$ in the upper (lower) trajectories. The transition (thick line) is crossed by varying the intensity of the speckle potential $V_0$ as shown on the figure. The red boxes on the upper (red) trajectories mark where the flow should be terminated for condensates of total length $20,100$ and $500~\mu m$.}
 \label{fig:RGflow}
\end{figure}

The flow proceeds along the trajectories
 \be
\label{eqn:RGFlowTrajectories}
f_0 = \alpha-\text{ln}\left[\a +1\right]+\epsilon
\ee
 shown in Fig. \ref{fig:RGflow}, where $\epsilon$ labels the trajectory. The initial conditions $f_0(\Gamma=0)$ and $\a(\Gamma=0)$ of the flow were derived above from the microscopic potential.
The phase transition is crossed by changing the chemical potential or the disorder strength, which moves the initial point across the separatrix $\epsilon=0$.

% Note that $\a$ is renormalized downward due to the fluctuations induced by the charging energies. While at the transition point the renormalized value of $\a$ is exactly zero, the bare value of the exponent at that point must be $\a_c>0$. For reasonably small values of the scattering length the typical charging energy is much smaller than the typical Josephson energy $J_0$. The system is then very close to the "classical" fixed line found in Ref. \onlinecite{AKPR} already at the outset and there is not much room for $\a$ to flow. Hence we expect the transition to occur near the point $\a=0$.

In Fig. \ref{fig:RGflow} we demonstrate the RG flow derived for two realistic setups of $^{87}$Rb condensates with disorder produced by a random speckle potential similar to Ref. \onlinecite{Billy2008}. Trap parameters are detailed in the figure caption and the disorder correlation length is taken to be the speckle size $\sigma \approx 0.26 \mu m$. The system is in the rough disorder regime in both cases. Experiments with atom-chip traps are in the opposite, smooth disorder, regime\cite{KrugerDisorder}. But as we have shown, this leads to the same universal behavior on large scales.

We also note that the specific structure of speckle potentials, non-gaussianity and spatial correlations, has known implications on Anderson localization. For example, emergence of
pseudo mobility edges\cite{Lugan2009,Gurevich2009}. However the exact structure of the speckle potential does not significantly affect the calculation of the Josephson elements. First, the Josephson elements are induced by tunneling of states under a long barrier and do not involve high energy states near the effective mobility edges. Second the exact probability distribution of $V$ and in particular the fact that it is asymmetric does not enter the calculation in any important way. For our general framework to be applicable, the probability distribution of $V$ should decay faster than  $e^{-\sqrt{V/V_0}}$ at large positive $V$. If the decay is slower, then the weak links are not determined by rare long barriers but rather by the less rare high barriers. Using the exact structure of speckle potential in place of the Gaussian potentials used here, will result in a small change of the computed parameters $\a$ and $f_0$, but not to a change in the form of the distributions.

We can now also address the finite size of the system. For condensates of increasing length the RG flow should be terminated at decreasing energy scales, i.e. further along the flow, when all the elements in the effective Josephson array have been eliminated. The red squares on the flow trajectories in Fig. \ref{fig:RGflow} demonstrate termination points corresponding to condensates of length $20, 100$ and $500 \mu m$. The calculated values of physical observables, such as the superfluid stiffness or the compressibility\cite{AKPR3}, should be recorded at the termination points and compared to the experimental measurements.
To further characterize the critical point it would be interesting to study the coherence properties of the disordered condensate that can be extracted from interference experiments\cite{Leanhardt2003}. This will require a generalization of the theory of fringe statistics\cite{AltmanPNAS,AltmanNP}
to the case of disordered condensates.

\section{Phase Diagram}

Above we have established a direct link between the microscopic model of bosons in a disordered potential and the renormalization group flow of the random Josephson array at strong disorder. This
now allows us to compute the phase diagram in the space of microscopic parameters, the dimensionless interaction and the disorder strength, at strong disorder. The transition line in that space is given by the set of models that map onto points on the separatrix as initial conditions for the RG flow in Fig. \ref{fig:RGflow}.

The separatrix is given in terms of the RG flow parameters by Eq. (\ref{eqn:RGFlowTrajectories})  with $\e=0$. We replace $f_0$ by its value as a function of the microscopic parameters (\ref{U-distribution}) to obtain
\be
\label{eq:phase_boundary}
\a (\hat{\mu})-\text{ln}\left[\a(\hat{\mu}) +1\right]=-\frac{4}{3} \hat{u}^2\ln(1-q_{\hat{\mu}}).
\ee
where $\hat{\mu} = \mu/E_*$ and $\hat{u} = u/E_* \zeta_*$. The right hand side of the equation encodes the effect of quantum fluctuations, induced by the charging energies, which cause the bending of the separatrix to positive values of $\alpha$. By contrast, the mean field stiffness vanishes only for $\a\le 0$.

The difference between the actual transition and the mean field approximation is most apparent when we plot the phase boundary in the space $\hat{\mu}$ versus $\hat{u}$. In the regime of rough disorder, the disorder strength enters only through the energy $E_*$. Therefore in this limit Eq. (\ref{eq:phase_boundary}) charts a universal phase boundary in the space $(\hat{\mu},\hat{u})$. We can change $\mu$ independently of the interaction $u$ by tuning the density. However in the classical (Gross-Pitaevskii) solution the interaction enters only through the chemical potential, and so within this approximation the system becomes insulating below a critical chemical potential $\hat{\mu}_*\approx 0.47$ independent of $\hat{u}$. On the other hand the condition
(\ref{eq:phase_boundary}) gives the transition line
\be
\hat{\mu}(\hat{u})=\hat{\mu}_*+0.89 \hat{u} -0.54 \hat{u}^2 + O\left(\hat{u}^3\right)
 \label{eq:mu_boundary}
 \ee
 showing a non trivial dependence on $\hat{u}$ as a correction to the mean field result. The system is insulating for chemical potentials below the transition line. To obtain the formula  (\ref{eq:mu_boundary}) we have expanded both sides of Eq. (\ref{eq:phase_boundary}) to
quadratic order in $\hat{\mu}-\hat{\mu}_*$.

We stress that the quantum corrections stem from quantum phase slips, generated by the charging term in the effective hamiltonian (\ref{eq:JJA_hamiltonian}). The RG flow accounts for such phase slips through decimation of sites with large charging energies, which is accompanied by renormalization of the local Josephson couplings. By contrast Bogoliubov theory, being a quadratic expansion around the GP solution, cannot give rise to renormalization of the stiffness and therefore does not result in a correction to the mean field phase diagram\cite{Fontanesi2011}.

%A phase diagram has been previously derived in this regime within the Gross-Pitaevskii approach which gives the mean field transition line\cite{Fontanesi2011}. A phase diagram charted using arguments of competition between the local stiffness and local charging energies gives essentially the same result as the mean field result.

%The real-space RG provides a criterion for the phase of the system. If the initial distributions are such that the point $(f_0,\a)$ is on the right of the separatrix in Fig. \ref{fig:RGflow}, the system is a superfluid. Otherwise, it is insulating. This enables us to find the phase diagram of the system. The diagrams are computed for the rough-disorder limit. We compare our results with the phase diagrams obtained by mean-field\cite{Fontanesi2010}, the analysis of Falco \emph{et al.}\cite{Falco2009} and the results obtained by Giamarchi and Schultz

%This prediction can be tested in experiments where the interaction $u$ is position dependent (e.g. by a non-homogenous magnetic field along the system tuned near the Freschbach resonance of the trapped atoms resulting with a position dependent scattering length). In this case, for chemical potential larger than $\hat{\mu}_*$, we predict that there will be regions in the system with small $u$ which are superfluid while other regions with larger $u$ will be insulating. The coexistence of the phases does not exist in the mean-field prediction.

We can now find the phase diagram in the space of disorder strength versus interaction.
The disorder in the rough limit is parameterized by $D=V_0^2\sigma=E_*^2 \zeta_*$. To obtain a universal phase diagram (i.e. independent of density and particle species) we turn to dimensionless parameters
normalizing energies by the degeneracy scale $T_d=\hbar^2 \rho^2 /2m$ and length scales by the inverse density. Hence we use $\tilde{D}=D \rho / T_d^2$ and $\tilde{u} = u \rho / T_d$.

As a first step we express Eq. (\ref{eq:mu_boundary}) as a condition on the density rather than the chemical potential by using the equation of state (\ref{EOS})
\begin{equation}
\label{eq:rho_boundary}
\hat{\rho}\approx 0.49/\hat{u} +1.00-0.54 \hat{u} + O\left(\hat{u}^2\right)
\end{equation}
where $\hat{\rho} = \rho \zeta_*$.
Finally, noting that $\tilde{u} = \hat{u}/\hat{\rho}$ and $\tilde{D} = 1/\hat{\rho}^3$ we can obtain the phase boundary in terms of $\tilde{u}$ and $\tilde{D}$ from (\ref{eq:rho_boundary}) using a simple change of variables. This gives the asymptotic phase boundary for weak interactions
\be
\label{eqn:D}
    \sqrt{\tilde{D}} \approx 1.70 \tilde{u}^{3/4}-1.81 \tilde{u}^{5/4}+O\left(\tilde{u}^{7/4}\right).
\ee

\begin{figure}
 \centerline{\resizebox{0.9\linewidth}{!}{\includegraphics{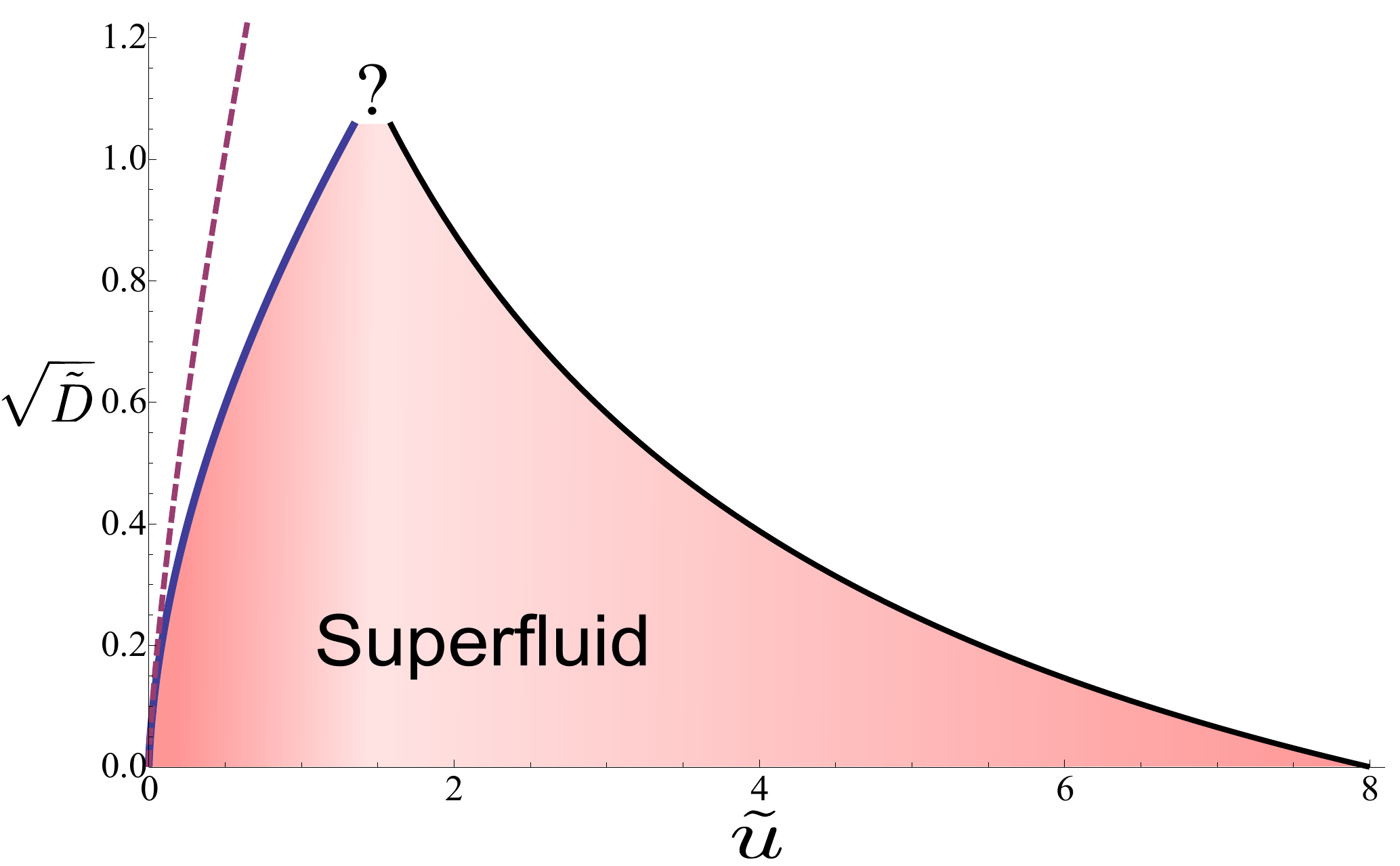}}}
 \caption{{\em $\sqrt{D}$-$u$ phase diagram.} The diagram is given in terms of dimensionless disorder strength $\sqrt{\tilde{D}}= \sqrt{D\rho}/T_d$ and the dimensionless interaction strength $\tilde{u}=u \rho/T_d$, where $\rho$ is the density, $u$ is the effective contact interaction, $D$ is the measure of rough disorder defined in the text, and $T_d=\hbar^2 \rho^2 /2m$ is the temperature of quantum degeneracy. The phase boundary at weak interactions corresponds to the set of points mapped to the separatrix of the RG flow. The dashed line above the true transition marks the points where the classical theory predicts a transition\cite{Fontanesi2010} based on vanishing of the mean field stiffness. The phase boundary at strong interactions is obtained from the weak disorder theory of Giamarchi and Schultz\cite{Giamarchi1988}(GS). }
 \label{fig:phase_diagram_D_u}
\end{figure}

In order to obtain the phase boundary at larger interaction strengths we take the exact form of the separatrix from Eq. \eqref{eq:phase_boundary} and use the exact equation of state \eqref{EOS} and solve the equation numerically. The phase boundary obtained in this way is shown in Fig. \ref{fig:phase_diagram_D_u}. It has the asymptotic form given by Eq. \eqref{eqn:D} at weak interactions. The mean field criterion $\a=0$, without the quantum corrections, gives only the first term in the expansion (\ref{eqn:D}), i.e. $\sqrt{\tilde{D}_{MF}}\approx 1.70 \tilde{u}^{3/4}$. This mean field transition line, previously found
in Refs. [\onlinecite{Fontanesi2010}] and [\onlinecite{Falco2009}] is also plotted on Fig. \ref{fig:phase_diagram_D_u} for comparison with the actual transition.

The phase boundary calculated using the mapping to a random Josephson array cannot be continued to arbitrarily strong interactions.
Moving on the line given by Eq. (\ref{eqn:D}) toward stronger interactions corresponds to moving the initial point of the RG flow on the separatrix shown in Fig. \ref{fig:RGflow} from the point $\a=f_0=0$ toward higher values of $\a$. This is tantamount to decreasing the disorder on the effective Josephson array, which makes the starting point of the RG flow gradually less controlled. Specifically, at the dimensionless interaction strength $\tilde{u}\gtrsim 1.5$ the exponent $\a$ reaches the value $\a\sim 1$. This can serve as a characteristic value beyond which the distribution of Josephson links becomes narrow and where the real space RG approach is not justified a priori.

To complement the phase diagram found in the weak interaction limit, we also obtain the phase boundary between the superfluid and the insulator in the regime of weak disorder. This is done using the perturbative RG flow of Giamarchi and Schulz\cite{Giamarchi1988}
\be
\label{eq:GS-flow}
\begin{split}
\frac{dK}{dl} =& \frac{1}{2} \tilde{D}\\
\frac{d\tilde{D}}{dl} =& \left( 3-2K^{-1}\right) \tilde{D}\\
\end{split}
\ee
Here we used the convention of Ref. [\onlinecite{Giamarchi1988}] for the Luttinger parameter $K$.
By the dividing the two equations, the separatrix of the flow is easily found to be given by $\tilde{D}= 6(K-K_0)-4\ln (K/K_0)$, where $K_0=2/3$ is the value of the Luttinger parameter at the fixed point.
Now, using the relation of the bare Luttinger parameter (i.e. $K$ of the clean system) to the microscopic interaction in the relevant regime $K^{-1}\approx 1+4/\tilde{u}$, \cite{Cazalilla2004} we  obtain the approximate phase boundary
$\sqrt{\tilde{D}}\approx (8-\tilde{u})/12\sqrt{2}$. Of course this approach is justified only as long as the disorder strength $\tilde{D}$ is smaller than $1$.
The global phase diagram inferred from both the strong and weak disorder limits is plotted in Fig. \ref{fig:phase_diagram_D_u}. % It remains an open question how to interpolate between the phase transitions found in the two regimes.

\begin{comment}
\begin{figure}
 \centerline{\resizebox{0.9\linewidth}{!}{\includegraphics{phase_diagram_mu_u_const_disorder.pdf}}}
 \caption{{\em $\mu$-$u$ phase diagram.} The phase diagram in the $\mu$-$u$ plane for constant disorder strength. The upper plot is the phase separatrix obtained by our method and the lower plot is the mean-field prediction, which does not depend on $u$}
 \label{fig:phase_diagram_mu_u}
\end{figure}
\end{comment}

\section{Conclusions} In this paper we derived an effective quantum Josephson array model starting from a realistic microscopic model of bosons in a one dimensional random potential. The distributions (\ref{PJ}) and (\ref{U-distribution}) of the effective coupling constants obtained in this way are precisely the stable solutions of the real-space RG found in Ref. \onlinecite{AKPR}. Their flow is determined by a strong randomness fixed point, which controls a quantum phase transition between a superfluid and insulating phase.

The ab-initio mapping from microscopic models of bosons in a random potential to precise initial conditions for the RG flow allows to make quantitative predictions and thereby can facilitate experimental detection
of the new critical point. This can be done for example by measuring the finite size scaling behavior of the phase correlations in interference experiments. The results can be directly checked for consistency with the finite size scaling implied by the RG flow near the critical point.

Bridging the gap between the microscopic physics and the RG flow of the Josephson array model allowed us to predict a phase diagram (Fig. \ref{fig:phase_diagram_D_u}) in the space of microscopic parameters, disorder and interaction strength, which is valid in the regime of weak interactions. The transition line is modified compared to that previously inferred from mean field (Gross-Pitaevskii) theory\cite{Fontanesi2009,Fontanesi2010} and from estimates based on typical values of the Josephson coupling compared to interaction strength\cite{Falco2009}.

Note that the critical point in this regime is different in nature from the weak disorder transition considered by Giamarchi and Schulz\cite{Giamarchi1988}, shown as a separate transition line in the strong interaction range of Fig. \ref{fig:phase_diagram_D_u}. At the strong randomness fixed point rare events in the form of weak links that effectively cut the chain play a central role, while these are completely neglected in the weak disorder theory. As shown in Ref. [\onlinecite{AKPR3}], such events lead to a transition at a non universal value of the phase-correlation decay exponent that is always smaller than the universal value of the exponent predicted by the weak disorder theory. It remains an interesting open question how the transition interpolates between the two limits.

\emph{Acknowledgements.} We thank Anatoli Polkovnikov, Yariv Kafri and David Huse for stimulating discussions.
This work was supported in part by NSF under grant No. PHY05-51164, the US Israel BSF, ISF, and a grant from the estate of Ernst and Anni Deutsch.

\appendix

%In the appendix we provide more details of the derivation of three important formulas
%used in the text. First, in sec. \ref{sec:stiffness} we obtain the expression given in Eq. (2) of the text for the phase stiffness of a Gross-Pitaevskii (GP) ground state in a random potential. In sec. \ref{sec:J} we outline the derivation of Eq. (3) of the text, which gives the Josephson coupling between two superfluid puddles. Finally, in sec. \ref{sec:PJ} we provide a rigorous derivation of the distribution of the Josephson coupling constants, confirming the result of the more heuristic treatment given in the main text.

\setcounter{secnumdepth}{2}
\section{Superfluid Stiffness}\label{sec:stiffness}
The Gross-Pitaevskii energy functional of the wave-function  $\psi(x)=\sqrt{\rho(x)}e^{i\f(x)}$ can be written as:
\be
\label{GP-Energy}
\begin{split}
E_\Phi[\rho,\f] &=\int_{0}^{L} dx \left({\frac{\rho}{2m}}(\partial_x\f)^2-\lambda\int_0^L dx\partial_x\f\right)\\
&\quad+\int_0^L dx\left(\frac{(\partial_x\rho)^2}{ 4m\rho}+V(x)\rho+\half u\rho^2\right)
\end{split}
\ee
where $\lambda$ is a Lagrange multiplier that can be used used to enforce a phase twist $\Phi=\int_0^L\partial_x\f$ along the length of the condensate . For $\Phi=0$ the ground state has uniform phase and can be chosen to be real $\psi_0=\sqrt{\rho_0}$. The energy of this state is $\e_0=E_0[\rho_0]$.

The superfluid stiffness is proportional to
the quadratic change of the energy with a small twist $\Phi$
\begin{equation}
\label{sf-def}
\rho_s = L \left(\frac{\partial^2 E}{\partial \Phi^2}\right)_{\Phi=0}.
\end{equation}
To compute $\rho_s$ we write the ground state wave-function in presence of
the twist as
$\psi_\Phi(x)=\sqrt{\rho_0+\d\rho(x)}e^{i\f(x)}$. and expand the energy to quadratic order in the changes $\d\rho$ and $\partial_x\f$
\be
E_\Phi=E_0+O(\d\rho^2) + \int_0^L dx\left({\rho_0\over 2m}(\partial_x\f)^2-\lambda\partial_x\f\right).
\label{EPhi}
\ee
We now note that the phase twist $\Phi$ led to a proportional phase gradient $\partial_x\f$, while the change in the local density must be quadratic $\d\rho\propto \Phi^2$. This is because $\rho$ is even under time-reversal whereas $\Phi$ is odd. Therefore the change in energy due to distortion of the density by the twist is
proportional to $\Phi^4$ and does not contribute to the stiffness.

Minimizing (\ref{EPhi}) with respect to the phase gradient we have
\begin{equation}
\label{grad-phi}
\partial_x \varphi(x) = - \frac{m}{\hbar^2} \frac{\lambda}{|\psi_0|^2}
\end{equation}
and $\lambda = - \frac{\hbar^2}{m} \frac{\Phi}{\int_0^L|\psi_0|^{-2}dx}$ is obtained by imposing the constraint. By substituting back in Eq. \eqref{EPhi} we finally obtain the superfluid stiffness
\begin{equation}
\label{sf-stiffness-result}
 \rho_s = L \frac{\hbar^2}{m} \frac{1}{\int_0^L |\psi_0|^{-2} dx}.
\end{equation}

\section{Josephson Coupling}\label{sec:J}
Using the above equation for the SF-stiffness, we may calculate the effective Josephson coupling of two neighboring SF puddles. The energy of the coupled puddles is $E\sim J cos(\Phi_1-\Phi_2)$. Therefore, it can be related to the SF stiffness of the system in the region between the two puddles by
 \begin{equation}
 \label{J-calculation}
 J_{12}=\rho_s/L=\frac{\hbar^2}{m} \frac{1}{\int_{x_1}^{x_2} |\psi_0|^{-2} dx},
  \end{equation}
  where $x_1$ ($x_2$) is the left (right) edge of the barrier between the two SF puddles. We are left with the problem of finding the wave-function in the region between the two puddles. This is done in a way similar to the calculation in Ref. \onlinecite{Huhtamaki}.

 since the wave-function amplitude is small under the barrier, the interaction term
 $u|\psi|^2$ is negligible. The wavefunction follows a linear Schr\"odinger equation
 and it can be approximated using the WKB approximation
\begin{equation}
\label{psi-WKB}
\psi_0(x) = \frac{C_1}{\sqrt{\kappa(x)}} e^{-\int_{x_1}^x dy \kappa(y)} + \frac{C_2}{\sqrt{\kappa(x)}} e^{\int_{x_2}^x dy \kappa(y)}
\end{equation}
with $\kappa(x) = \sqrt{\frac{2m}{\hbar^2} \left[ \mu - V(x) \right]}$ and $\mu$ the chemical potential. The effect of the puddles enters through the pre-factors $C_1$ and $C_2$. These are determined by the matching conditions of the WKB wave-function with the wave-function inside the puddles.

In the regime where the healing length of the puddles $\xi_h$ is smaller than the size of the puddles, we may use the Thomas-Fermi (TF) wave-function inside the puddles $\psi_0(x)=\sqrt{\frac{\mu-V(x)}{u}}$. However, the TF-approximation breaks down at distances closer than $\sim\xi_h$ from the edge of the puddles. In this narrow region the potential can be taken to be linear, and since the amplitude is small
we can again drop the interaction term. The solutions to the Schr\"odinger equation in this region are Airy functions, which we can match with the WKB wavefunction and with the TF wave-functions on the two sides. This gives the constants:
\begin{equation}C_i = \sqrt{\frac{|V'(x_i)|}{2u}} e^{1/3} \frac{Ai \left[ 2^{-2/3}\right]}{Ai \left[ 2^{-2/3}\right]} \approx 0.4 \sqrt{\frac{|V'(x_i)|}{u}}.
\end{equation}

Having obtained the wave-function $\psi_0(x)$ we are in position to compute the superfluid stiffness using \eqref{J-calculation}. Because $\psi_0$ is exponentially suppressed in the middle of the barrier the integrand $|\psi_0|^{-2}$ is strongly peaked suggesting the use of a saddle point approximation to evaluate the integral. We write the integral as $\int dx |\psi_0(x)|^{-2}= \int dx \kappa (x) e^{f(x)}\approx \int dx \kappa (x_0) e^{f(x_0) +\frac{1}{2} f''(x_0)(x-x_0)^2}$, with the obvious definition of $f(x)$. We approximate it using a saddle point approximation around $x_0$ which is the position of the maximum of $f(x)$. By differentiating $f(x)$ one can show that $x_0$ satisfies $f''(x_0) = -2 \kappa^2(x_0)$ and $e^{-\frac{1}{2} f(x_0)}=2 \sqrt{C_1 C_2} e^{-\frac{1}{2} \int_{x_1}^{x_2}dx \kappa(x)}$. Using these in the saddle point approximation, the Josephson coupling is found by gaussian integration
\begin{equation}
\label{J-result}
J_{12} = \frac{4}{\sqrt{\pi}}\frac{ \hbar^2}{m} C_1 C_2 e^{-\int_{x_1}^{x_2} dx \kappa(x)}.
\end{equation}
$J$ is the product of the tunneling coefficient, denoted by $T$, and a non-trivial pre-facor.

\section{Distribution of the tunneling coefficient}\label{sec:PJ}

The tunneling coefficient $T$ is the exponential factor in eq. \eqref{J-result}
\be
\label{T-Def}
T = e^{-\int_{x_1}^{x_2} dx \sqrt{\frac{2m}{\hbar^2} \left[ \mu - V(x) \right]}}.
\ee
We shall approximate the integral in the exponent as a sum on segments of size $\sigma$, the correlation length of the potential, and take $V(x)=V_i$, a constant on each segment
\be
\begin{split}
I&=\int_{x_1}^{x_2} dx\sqrt{\frac{2m}{\hbar^2}(V(x)-\mu)}\\
&\approx\s\sum_{i=1}^l\sqrt{\frac{2m}{\hbar^2}(V_i-\mu)}\equiv \sum_{i=1}^l z_i.
\label{Isum}
\end{split}
\ee
The distribution of the potential in each segment is given by the conditional probability
\be
P_\mu(V)=P(V|V>\mu)={1\over q_\mu \sqrt{2\pi V_0^2}}e^{-V^2/2V_0^2},
\ee
where $q_\mu=P(V>\mu)$ is the integrated probability for the potential to be higher than the chemical potential $\mu$. The number of summands $l=(x_2-x_1)/\sigma$ is a random variable in itself with the distribution
\be
\label{l-discrete-dist}
p(l) = q_\mu^{l-1} (1-q_\mu)\approx |\text{ln} q_\mu| \exp (-| \text{ln} q_\mu|l)
\ee
In the last equality we have taken the continuum limit of $l$. This is not expected to affect the distribution of $T$ in the limit of small $T$, since this tail is controlled by large values of $l$.

We can now derive the probability distribution of the sum $I$. Formally it is given by
\be
\label{I-dist-1}
P(I) = \int_0^\infty dl p(I|l) p(l)
\ee
where $p(I|l)$ is the conditional probability of $I$ given a barrier length $l$. Since $I$ for a given $l$ is a sum of   independent and identically distributed random variables the characteristic function $\Phi_l$ is simply related to the characteristic function of each term $z_i$ as $\Phi_l(t) = \phi_z(t)^l$.
The conditional distribution $p(I|l)$ is given by the inverse fourier transform of the characteristic function
\be
\label{cond-prob-I}
p(I|l) = \frac{1}{2 \pi} \int_{-\infty}^{\infty} dt \exp (-i t I)  \phi_z(i t)^l
\ee
Putting together Eqs. (\ref{l-discrete-dist}), (\ref{I-dist-1}) and  (\ref{cond-prob-I}) we get
\begin{eqnarray*}
p(I) & = & \frac{1}{2 \pi}\int_0^\infty dl \int_{-\infty}^{\infty} dt e^{-i t I}  \phi_z(t)^l |\text{ln} q_\mu| e^{-|\text{ln} q_\mu|l} \\
&= &\frac{|\text{ln} q_\mu|}{2 \pi} \int_{-\infty}^{\infty} dt e^{-i t I} \int_0^\infty dl  e^{ -l\left(|\ln q_\mu|-\ln\phi_z(t)\right)}  \\
& = & \frac{|\text{ln} q_\mu|}{2 \pi} \int_{-\infty}^{\infty} dt e^{-i t I} \frac{1}{|\text{ln}q_\mu|-\text{ln}\left(\phi_z(t)\right)} \\
& = & \frac{|\text{ln} q_\mu|}{2 \pi i} \int_{-i \infty}^{i \infty} d\tau e^{-\tau I} \frac{1}{|\text{ln}q_\mu|-\text{ln}\left(\phi_z(-i\tau)\right)}.
\end{eqnarray*}
In the last line we changed variable $\tau=i t$, and integrated over the imaginary axis in the complex plane.

\begin{figure}
 \centerline{\resizebox{0.9\linewidth}{!}{\includegraphics{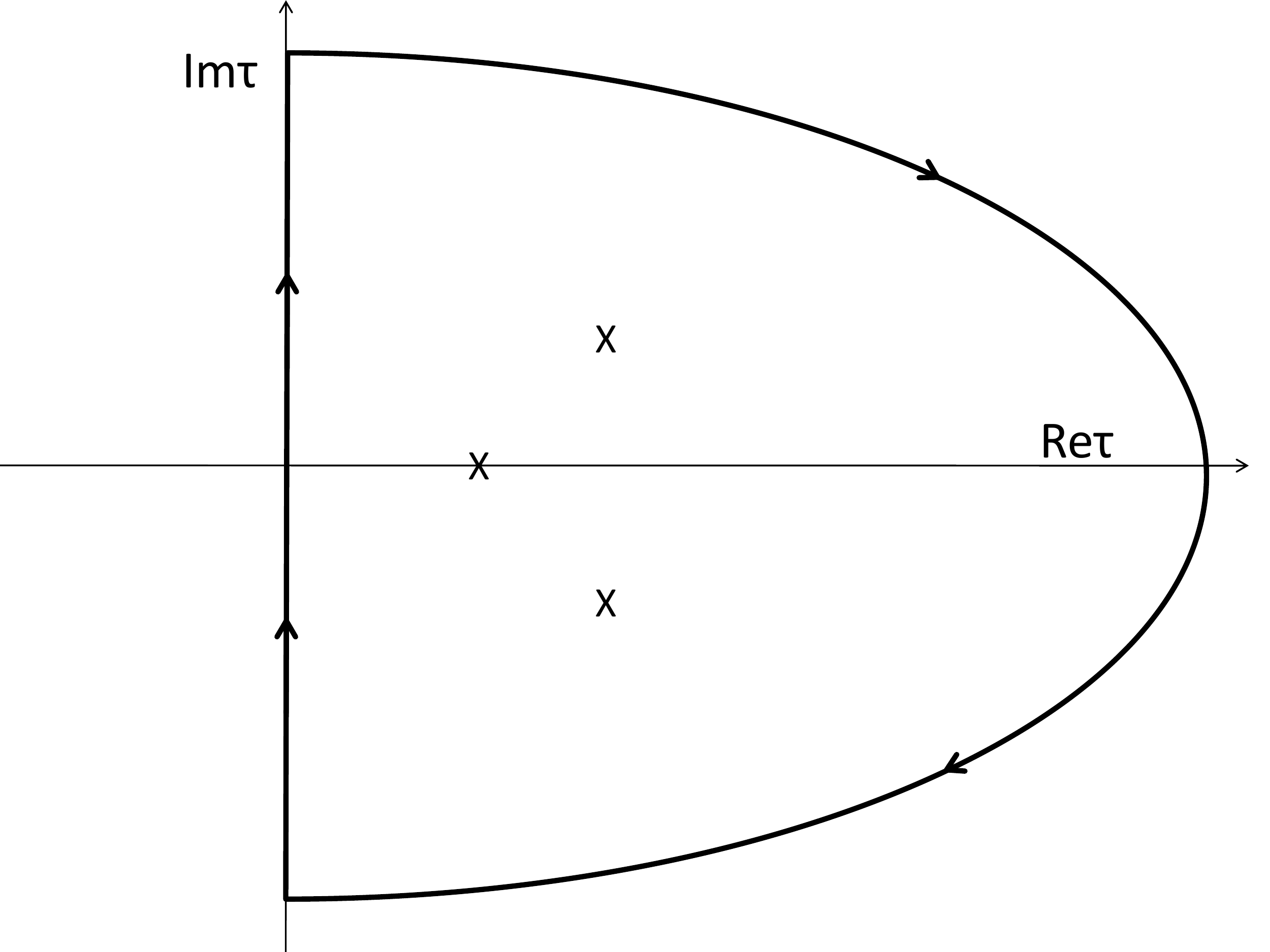}}}
 \caption{{\em The contour integration.} The contour integration in the complex plane over the right semi-circle. The integral collects all the poles in the right half plane}
 \label{fig:contour}
\end{figure}

 We can close the integration contour at infinity over the right semi-circle, where the integrand decays exponentially, without changing the integral (see Fig. \ref{fig:contour}). The contributions to the integral then come from the poles at the points $\tau_i$ having a positive real part
 \begin{equation}
p(I) = \sum_{i, Re(\tau_i)>0} e^{-\tau_i I} \text{Res}\left(\frac{1}{\text{ln}q_\mu+\text{ln}\left[\phi_z(-i\tau) \right]}\right)_{\tau_i}.
\end{equation}
 The points $\tau_i$ are zeros of the denominator and therefore obey the equation $\phi_z(-i\tau_i)=1/q_\mu$.

We shall now prove a theorem regarding the solutions of the above equation. This theorem will enable us to predict the tail of the distribution of $T$.
\newtheorem{zero_theorem}{theorem}
\begin{zero_theorem}
The pole condition $\phi_z(-i\tau)=1/q_\mu$ has a real solution $\tau_0>0$. All other solutions $\tau_i$ satisfy $Re \, \tau_i>\tau_0$.
\end{zero_theorem}
\begin{proof}
First, we prove the first part of the theorem. Using the definition of the characteristic functions with real argument $\tau$ we get the pole condition
\be
\label{pole-condition}
\int_0^{\infty} dz e^{\tau z} p(z)=1/q_\mu.
\ee
Note that the lower limit of the integration is zero and not $-\infty$ since $p(z<0)=0$. $q_\mu$ is a probability, hence the R.H.S. of \eqref{pole-condition} is larger than $1$. The integral on the L.H.S. converge for any real $\tau$ since $p(z)$ decays faster then any exponential. By differentiation one can show that this integral is a monotonically increasing smooth function of $\tau$. For $\tau=0$ it is $1$. Therefore, there is always some real $\tau_0>0$ which satisfies the pole condition.

We prove the second part of the theorem by contradiction. Let us assume that there is a {\em complex} pole at $\tau_i = a+ i b$ with $a\leq \tau_0$. Both $\tau_i$ and $\tau_0$ obey the pole condition, and so
$\phi_z(-i\tau_0)=1/q_\mu=\phi_z(-i\tau_i)$. Using the definition of the characteristic functions this equality can be written as:
\be
\begin{split}
0&=\int_0^{\infty} dz \left(e^{\tau_0 z}-e^{\tau_i z}\right)p(z) \\
&=\int_0^{\infty} dz e^{a z}\left(e^{(\tau_0 - a) z}-e^{i b z}\right)p(z)
\end{split}
\ee
Now, by our assumption that $a\leq\tau$ the real part of the expressions in the brackets above is positive (it may be zero on a discrete set of points if $a=\tau_0$). Since all other factors in the integrand are non-negative, we conclude that the integral cannot vanish, which contradicts our assumption. Therefore we must have $a=Re\, \tau_i>\tau_0$.
\end{proof}

The distribution of $T$ as $T\to 0$ is determined by the tail of the distribution $p(I)$ and it is therefore dominated by the pole $\tau_0$ with smallest real part. This pole must be real according to the above theorem. It follows, by change of variables, that the distribution of $T$ behaves as a power-law in the limit of small $T$. That is, $P(T)=A T^{\tau_0-1}$, where $A$ is the residue of the pole at $\tau_0$.

To obtain the numerical value of the exponent we have to find a real solution to the pole condition. This is done by expanding the log of the characteristic function in powers of $\tau$
\be
\text{ln} \left[\phi_i(\tau) \right]= \sum_{n=1}^{\infty} \frac{\tau^n}{n!} \kappa_n,
\ee
where $\kappa_n$ is the $n^\text{th}$ cumulant of the distribution $p(z_i)$. The first two cumulants are $\kappa_1 = \langle z_i \rangle = \bar{z_i}$ and $\kappa_2 = \langle z_i^2 \rangle- \langle z_i \rangle^2=\delta z_i^2$. In general, the expansion has some finite radius of convergence. We will come back to this point later.

Plugging the expansion in the pole condition $\phi(-i\tau)=1/q_\mu$ gives the infinite series equation
\be
\label{cumulant-eq}
\sum_{n=1}^{\infty} \frac{\tau^n}{n!} \kappa_n + \text{ln}q_\mu = 0.
\ee
To approximate the real zero we truncate the series and find the zeros of the resulting polynomial. The zeros of the truncated polynomials have the following properties\cite{Christiansen2006}:
\begin{enumerate}
\item If the zero is within the radius of convergence, the truncated polynomial will converge to the true zero as the length of the polynomial is increased.
\item If the zero is outside the radius of convergence, it will not be found.
\item Zeros of the polynomial which are not really zeros of the infinite series will cluster along the radius of convergence. We call these zeros spurious zeros.
\end{enumerate}

\begin{figure}
 \centerline{\resizebox{0.9\linewidth}{!}{\includegraphics{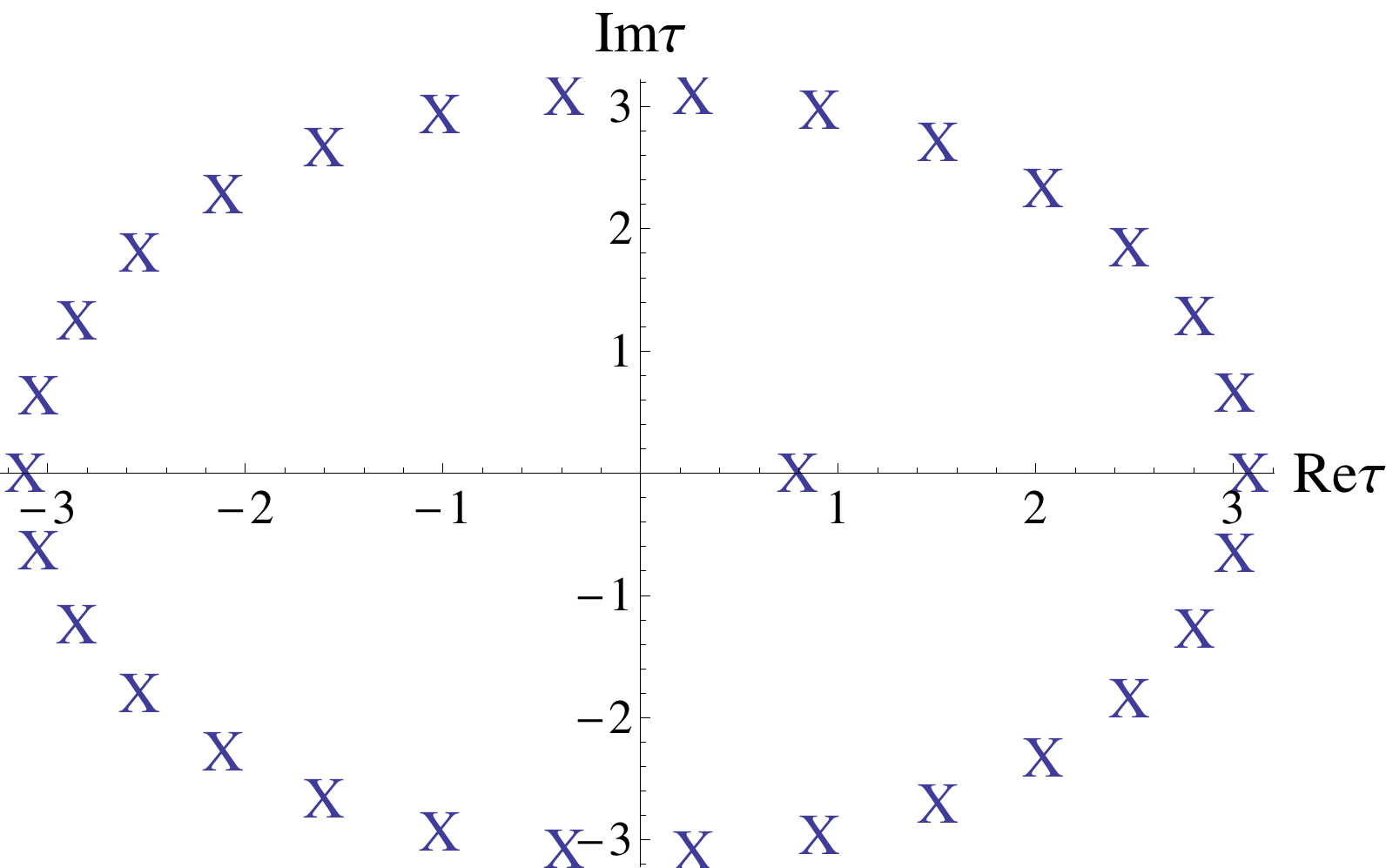}}}
 \caption{{\em Zeros of the truncated polynomial.} The zeros of the truncated polynomial ($31^{\text{st}}$ order) in eq. \eqref{cumulant-eq} which correspond to poles in the contour integration, as calculated for $\mu=0$. Spurious zeros cluster along a circle, which is the radius of convergence of the cumulant expansion. The real zero is inside the circle.}
 \label{fig:poles}
\end{figure}

Since the real zero that we look for is generally within the radius of convergence, we may truncate the series to approximate its value. We keep only the first two terms of the cumulant expansion, which gives us a $2^{\text{nd}}$ order polynomial, with the positive solution
\be
\label{tau0-result}
\tau_0 \approx  \frac{-\bar{z_i}+\sqrt{\bar{z_i}-2 \delta z_i^2 \text{ln}q_\mu}}{\delta z_i^2},
\ee
and the residue, which is also the pre-factor of the power-law tail
\be
\text{Res}\left(\frac{1}{|\text{ln}q_\mu|-\text{ln}\left(\phi_i(\tau) \right)}\right)_{\tau_0} \approx -\text{ln}q_\mu \frac{1}{\sqrt{\bar{z_i}-2 \delta z_i^2 \text{ln}q_\mu}}
\ee

Taking higher orders of the polynomial changes the position of the real pole only slightly. In fact $\tau_0$ is approximated by \eqref{tau0-result} with less than $0.03\%$ error.
All other zeros of the polynomial cluster along the radius of convergence, as expected. This is demonstrated for expansion up to $31^{\text{st}}$ order in fig. \ref{fig:poles} for $\mu=0$.

\bibliographystyle{phd-url-notitle}
\bibliography{disorder}

\begin{thebibliography}{28}
\expandafter\ifx\csname natexlab\endcsname\relax\def\natexlab#1{#1}\fi
\expandafter\ifx\csname bibnamefont\endcsname\relax
  \def\bibnamefont#1{#1}\fi
\expandafter\ifx\csname bibfnamefont\endcsname\relax
  \def\bibfnamefont#1{#1}\fi
\expandafter\ifx\csname citenamefont\endcsname\relax
  \def\citenamefont#1{#1}\fi
\expandafter\ifx\csname url\endcsname\relax
  \def\url#1{\texttt{#1}}\fi
\expandafter\ifx\csname urlprefix\endcsname\relax\def\urlprefix{URL }\fi
\providecommand{\bibinfo}[2]{#2}
\providecommand{\eprint}[2][]{\url{#2}}

\bibitem[{\citenamefont{Billy et~al.}(2008)\citenamefont{Billy, Josse, Zuo,
  Bernard, Hambrecht, Lugan, Cl\'{e}ment, Sanchez-Palencia, Bouyer, and
  Aspect}}]{Billy2008}
\bibinfo{author}{\bibfnamefont{J.}~\bibnamefont{Billy}},
  \bibinfo{author}{\bibfnamefont{V.}~\bibnamefont{Josse}},
  \bibinfo{author}{\bibfnamefont{Z.}~\bibnamefont{Zuo}},
  \bibinfo{author}{\bibfnamefont{A.}~\bibnamefont{Bernard}},
  \bibinfo{author}{\bibfnamefont{B.}~\bibnamefont{Hambrecht}},
  \bibinfo{author}{\bibfnamefont{P.}~\bibnamefont{Lugan}},
  \bibinfo{author}{\bibfnamefont{D.}~\bibnamefont{Cl\'{e}ment}},
  \bibinfo{author}{\bibfnamefont{L.}~\bibnamefont{Sanchez-Palencia}},
  \bibinfo{author}{\bibfnamefont{P.}~\bibnamefont{Bouyer}}, \bibnamefont{and}
  \bibinfo{author}{\bibfnamefont{A.}~\bibnamefont{Aspect}},
  \bibinfo{journal}{Nature} \textbf{\bibinfo{volume}{453}},
  \bibinfo{pages}{891} (\bibinfo{year}{2008}), ISSN \bibinfo{issn}{1476-4687},
  \href{http://www.ncbi.nlm.nih.gov/pubmed/18548065}{URL}.

\bibitem[{\citenamefont{Lye et~al.}(2005)\citenamefont{Lye, Fallani, Modugno,
  Wiersma, Fort, and Inguscio}}]{Inguscio}
\bibinfo{author}{\bibfnamefont{J.~E.} \bibnamefont{Lye}},
  \bibinfo{author}{\bibfnamefont{L.}~\bibnamefont{Fallani}},
  \bibinfo{author}{\bibfnamefont{M.}~\bibnamefont{Modugno}},
  \bibinfo{author}{\bibfnamefont{D.~S.} \bibnamefont{Wiersma}},
  \bibinfo{author}{\bibfnamefont{C.}~\bibnamefont{Fort}}, \bibnamefont{and}
  \bibinfo{author}{\bibfnamefont{M.}~\bibnamefont{Inguscio}},
  \bibinfo{journal}{Phys. Rev. Lett.} \textbf{\bibinfo{volume}{95}},
  \bibinfo{pages}{070401} (\bibinfo{year}{2005}),
  \href{http://link.aps.org/doi/10.1103/PhysRevLett.95.070401}{URL}.

\bibitem[{\citenamefont{Chen et~al.}(2008)\citenamefont{Chen, Hitchcock, Dries,
  Junker, Welford, and Hulet}}]{Chen2008a}
\bibinfo{author}{\bibfnamefont{Y.~P.} \bibnamefont{Chen}},
  \bibinfo{author}{\bibfnamefont{J.}~\bibnamefont{Hitchcock}},
  \bibinfo{author}{\bibfnamefont{D.}~\bibnamefont{Dries}},
  \bibinfo{author}{\bibfnamefont{M.}~\bibnamefont{Junker}},
  \bibinfo{author}{\bibfnamefont{C.}~\bibnamefont{Welford}}, \bibnamefont{and}
  \bibinfo{author}{\bibfnamefont{R.~G.} \bibnamefont{Hulet}},
  \bibinfo{journal}{Phys. Rev. A} \textbf{\bibinfo{volume}{77}},
  \bibinfo{pages}{033632} (\bibinfo{year}{2008}),
  \href{http://link.aps.org/doi/10.1103/PhysRevA.77.033632}{URL}.

\bibitem[{\citenamefont{Altman et~al.}(2004)\citenamefont{Altman, Kafri,
  Polkovnikov, and Refael}}]{AKPR}
\bibinfo{author}{\bibfnamefont{E.}~\bibnamefont{Altman}},
  \bibinfo{author}{\bibfnamefont{Y.}~\bibnamefont{Kafri}},
  \bibinfo{author}{\bibfnamefont{A.}~\bibnamefont{Polkovnikov}},
  \bibnamefont{and} \bibinfo{author}{\bibfnamefont{G.}~\bibnamefont{Refael}},
  \bibinfo{journal}{Phys. Rev. Lett.} \textbf{\bibinfo{volume}{93}},
  \bibinfo{pages}{150402} (\bibinfo{year}{2004}).

\bibitem[{\citenamefont{Lugan et~al.}(2007)\citenamefont{Lugan, Cl\'ement,
  Bouyer, Aspect, Lewenstein, and Sanchez-Palencia}}]{Lugan2007}
\bibinfo{author}{\bibfnamefont{P.}~\bibnamefont{Lugan}},
  \bibinfo{author}{\bibfnamefont{D.}~\bibnamefont{Cl\'ement}},
  \bibinfo{author}{\bibfnamefont{P.}~\bibnamefont{Bouyer}},
  \bibinfo{author}{\bibfnamefont{A.}~\bibnamefont{Aspect}},
  \bibinfo{author}{\bibfnamefont{M.}~\bibnamefont{Lewenstein}},
  \bibnamefont{and}
  \bibinfo{author}{\bibfnamefont{L.}~\bibnamefont{Sanchez-Palencia}},
  \bibinfo{journal}{Phys. Rev. Lett.} \textbf{\bibinfo{volume}{98}},
  \bibinfo{pages}{170403} (\bibinfo{year}{2007}),
  \href{http://link.aps.org/doi/10.1103/PhysRevLett.98.170403}{URL}.

\bibitem[{\citenamefont{Falco et~al.}(2009)\citenamefont{Falco, Nattermann, and
  Pokrovsky}}]{Falco2009}
\bibinfo{author}{\bibfnamefont{G.~M.} \bibnamefont{Falco}},
  \bibinfo{author}{\bibfnamefont{T.}~\bibnamefont{Nattermann}},
  \bibnamefont{and} \bibinfo{author}{\bibfnamefont{V.~L.}
  \bibnamefont{Pokrovsky}}, \bibinfo{journal}{Phys. Rev. B}
  \textbf{\bibinfo{volume}{80}}, \bibinfo{pages}{104515}
  (\bibinfo{year}{2009}),
  \href{http://link.aps.org/doi/10.1103/PhysRevB.80.104515}{URL}.

\bibitem[{\citenamefont{Aleiner et~al.}(2010)\citenamefont{Aleiner, Altshuler,
  and Shlyapnikov}}]{Aleiner2010}
\bibinfo{author}{\bibfnamefont{I.~L.} \bibnamefont{Aleiner}},
  \bibinfo{author}{\bibfnamefont{B.~L.} \bibnamefont{Altshuler}},
  \bibnamefont{and} \bibinfo{author}{\bibfnamefont{G.~V.}
  \bibnamefont{Shlyapnikov}}, \bibinfo{journal}{Nature Physics}
  \textbf{\bibinfo{volume}{6}}, \bibinfo{pages}{900} (\bibinfo{year}{2010}),
  ISSN \bibinfo{issn}{1745-2473},
  \href{http://www.nature.com/doifinder/10.1038/nphys1758}{URL}.

\bibitem[{\citenamefont{Haldane}(1981)}]{Haldane1981}
\bibinfo{author}{\bibfnamefont{F.~D.~M.} \bibnamefont{Haldane}},
  \bibinfo{journal}{Phys. Rev. Lett.} \textbf{\bibinfo{volume}{47}},
  \bibinfo{pages}{1840} (\bibinfo{year}{1981}),
  \href{http://link.aps.org/doi/10.1103/PhysRevLett.47.1840}{URL}.

\bibitem[{\citenamefont{Giamarchi and Schulz}(1988)}]{Giamarchi1988}
\bibinfo{author}{\bibfnamefont{T.}~\bibnamefont{Giamarchi}} \bibnamefont{and}
  \bibinfo{author}{\bibfnamefont{H.~J.} \bibnamefont{Schulz}},
  \bibinfo{journal}{Phys. Rev. B} \textbf{\bibinfo{volume}{37}},
  \bibinfo{pages}{325} (\bibinfo{year}{1988}),
  \href{http://link.aps.org/doi/10.1103/PhysRevB.37.325}{URL}.

\bibitem[{\citenamefont{Altman et~al.}(2008)\citenamefont{Altman, Kafri,
  Polkovnikov, and Refael}}]{AKPR2}
\bibinfo{author}{\bibfnamefont{E.}~\bibnamefont{Altman}},
  \bibinfo{author}{\bibfnamefont{Y.}~\bibnamefont{Kafri}},
  \bibinfo{author}{\bibfnamefont{A.}~\bibnamefont{Polkovnikov}},
  \bibnamefont{and} \bibinfo{author}{\bibfnamefont{G.}~\bibnamefont{Refael}},
  \bibinfo{journal}{Phys. Rev. Lett.} \textbf{\bibinfo{volume}{100}},
  \bibinfo{pages}{170402} (\bibinfo{year}{2008}).

\bibitem[{\citenamefont{Altman et~al.}(2010)\citenamefont{Altman, Kafri,
  Polkovnikov, and Refael}}]{AKPR3}
\bibinfo{author}{\bibfnamefont{E.}~\bibnamefont{Altman}},
  \bibinfo{author}{\bibfnamefont{Y.}~\bibnamefont{Kafri}},
  \bibinfo{author}{\bibfnamefont{A.}~\bibnamefont{Polkovnikov}},
  \bibnamefont{and} \bibinfo{author}{\bibfnamefont{G.}~\bibnamefont{Refael}},
  \bibinfo{journal}{Phys. Rev. B} \textbf{\bibinfo{volume}{81}},
  \bibinfo{pages}{174528} (\bibinfo{year}{2010}).

\bibitem[{\citenamefont{Dasgupta and Ma}(1980)}]{Dasgupta1980}
\bibinfo{author}{\bibfnamefont{C.}~\bibnamefont{Dasgupta}} \bibnamefont{and}
  \bibinfo{author}{\bibfnamefont{S.-k.} \bibnamefont{Ma}},
  \bibinfo{journal}{Phys. Rev. B} \textbf{\bibinfo{volume}{22}},
  \bibinfo{pages}{1305} (\bibinfo{year}{1980}),
  \href{http://link.aps.org/doi/10.1103/PhysRevB.22.1305}{URL}.

\bibitem[{\citenamefont{Fisher}(1994)}]{Fisher1994}
\bibinfo{author}{\bibfnamefont{D.~S.} \bibnamefont{Fisher}},
  \bibinfo{journal}{Phys. Rev. B} \textbf{\bibinfo{volume}{50}},
  \bibinfo{pages}{3799} (\bibinfo{year}{1994}),
  \href{http://link.aps.org/doi/10.1103/PhysRevB.50.3799}{URL}.

\bibitem[{\citenamefont{Fontanesi et~al.}(2011)\citenamefont{Fontanesi,
  Wouters, and Savona}}]{Fontanesi2011}
\bibinfo{author}{\bibfnamefont{L.}~\bibnamefont{Fontanesi}},
  \bibinfo{author}{\bibfnamefont{M.}~\bibnamefont{Wouters}}, \bibnamefont{and}
  \bibinfo{author}{\bibfnamefont{V.}~\bibnamefont{Savona}},
  \bibinfo{journal}{Phys. Rev. A} \textbf{\bibinfo{volume}{83}},
  \bibinfo{pages}{033626} (\bibinfo{year}{2011}),
  \href{http://link.aps.org/doi/10.1103/PhysRevA.83.033626}{URL}.

\bibitem[{\citenamefont{Lifshitz}(1965)}]{Lifshitz1965}
\bibinfo{author}{\bibfnamefont{I.~M.} \bibnamefont{Lifshitz}},
  \bibinfo{journal}{Soviet Physics Uspekhi} \textbf{\bibinfo{volume}{7}},
  \bibinfo{pages}{549} (\bibinfo{year}{1965}), ISSN \bibinfo{issn}{0038-5670},
  \href{http://stacks.iop.org/0038-5670/7/i=4/a=R03?key=crossref.e9a11fcc53b361f1e6436ead950b487a}{URL}.

\bibitem[{\citenamefont{Halperin and Lax}(1966)}]{Halperin1966}
\bibinfo{author}{\bibfnamefont{B.}~\bibnamefont{Halperin}} \bibnamefont{and}
  \bibinfo{author}{\bibfnamefont{M.}~\bibnamefont{Lax}},
  \bibinfo{journal}{Physical Review} \textbf{\bibinfo{volume}{148}},
  \bibinfo{pages}{722} (\bibinfo{year}{1966}), ISSN \bibinfo{issn}{0031-899X},
  \href{http://link.aps.org/doi/10.1103/PhysRev.148.722}{URL}.

\bibitem[{\citenamefont{Sanchez-Palencia}(2006)}]{Sanchez-Palencia2006}
\bibinfo{author}{\bibfnamefont{L.}~\bibnamefont{Sanchez-Palencia}},
  \bibinfo{journal}{Phys. Rev. A} \textbf{\bibinfo{volume}{74}},
  \bibinfo{pages}{053625} (\bibinfo{year}{2006}),
  \href{http://link.aps.org/doi/10.1103/PhysRevA.74.053625}{URL}.

\bibitem[{\citenamefont{Kr\"uger et~al.}(2007)\citenamefont{Kr\"uger,
  Andersson, Wildermuth, Hofferberth, Haller, Aigner, Groth, Bar-Joseph, and
  Schmiedmayer}}]{KrugerDisorder}
\bibinfo{author}{\bibfnamefont{P.}~\bibnamefont{Kr\"uger}},
  \bibinfo{author}{\bibfnamefont{L.~M.} \bibnamefont{Andersson}},
  \bibinfo{author}{\bibfnamefont{S.}~\bibnamefont{Wildermuth}},
  \bibinfo{author}{\bibfnamefont{S.}~\bibnamefont{Hofferberth}},
  \bibinfo{author}{\bibfnamefont{E.}~\bibnamefont{Haller}},
  \bibinfo{author}{\bibfnamefont{S.}~\bibnamefont{Aigner}},
  \bibinfo{author}{\bibfnamefont{S.}~\bibnamefont{Groth}},
  \bibinfo{author}{\bibfnamefont{I.}~\bibnamefont{Bar-Joseph}},
  \bibnamefont{and}
  \bibinfo{author}{\bibfnamefont{J.}~\bibnamefont{Schmiedmayer}},
  \bibinfo{journal}{Phys. Rev. A} \textbf{\bibinfo{volume}{76}},
  \bibinfo{pages}{063621} (\bibinfo{year}{2007}),
  \href{http://link.aps.org/doi/10.1103/PhysRevA.76.063621}{URL}.

\bibitem[{\citenamefont{Lugan et~al.}(2009)\citenamefont{Lugan, Aspect,
  Sanchez-Palencia, Delande, Gr\'emaud, M\"uller, and Miniatura}}]{Lugan2009}
\bibinfo{author}{\bibfnamefont{P.}~\bibnamefont{Lugan}},
  \bibinfo{author}{\bibfnamefont{A.}~\bibnamefont{Aspect}},
  \bibinfo{author}{\bibfnamefont{L.}~\bibnamefont{Sanchez-Palencia}},
  \bibinfo{author}{\bibfnamefont{D.}~\bibnamefont{Delande}},
  \bibinfo{author}{\bibfnamefont{B.}~\bibnamefont{Gr\'emaud}},
  \bibinfo{author}{\bibfnamefont{C.~A.} \bibnamefont{M\"uller}},
  \bibnamefont{and}
  \bibinfo{author}{\bibfnamefont{C.}~\bibnamefont{Miniatura}},
  \bibinfo{journal}{Phys. Rev. A} \textbf{\bibinfo{volume}{80}},
  \bibinfo{pages}{023605} (\bibinfo{year}{2009}),
  \href{http://link.aps.org/doi/10.1103/PhysRevA.80.023605}{URL}.

\bibitem[{\citenamefont{Gurevich and Kenneth}(2009)}]{Gurevich2009}
\bibinfo{author}{\bibfnamefont{E.}~\bibnamefont{Gurevich}} \bibnamefont{and}
  \bibinfo{author}{\bibfnamefont{O.}~\bibnamefont{Kenneth}},
  \bibinfo{journal}{Phys. Rev. A} \textbf{\bibinfo{volume}{79}},
  \bibinfo{pages}{063617} (\bibinfo{year}{2009}),
  \href{http://link.aps.org/doi/10.1103/PhysRevA.79.063617}{URL}.

\bibitem[{\citenamefont{Leanhardt et~al.}(2003)\citenamefont{Leanhardt, Shin,
  Chikkatur, Kielpinski, Ketterle, and Pritchard}}]{Leanhardt2003}
\bibinfo{author}{\bibfnamefont{A.~E.} \bibnamefont{Leanhardt}},
  \bibinfo{author}{\bibfnamefont{Y.}~\bibnamefont{Shin}},
  \bibinfo{author}{\bibfnamefont{A.~P.} \bibnamefont{Chikkatur}},
  \bibinfo{author}{\bibfnamefont{D.}~\bibnamefont{Kielpinski}},
  \bibinfo{author}{\bibfnamefont{W.}~\bibnamefont{Ketterle}}, \bibnamefont{and}
  \bibinfo{author}{\bibfnamefont{D.~E.} \bibnamefont{Pritchard}},
  \bibinfo{journal}{Phys. Rev. Lett.} \textbf{\bibinfo{volume}{90}},
  \bibinfo{pages}{100404} (\bibinfo{year}{2003}),
  \href{http://link.aps.org/doi/10.1103/PhysRevLett.90.100404}{URL}.

\bibitem[{\citenamefont{Polkovnikov et~al.}(2006)\citenamefont{Polkovnikov,
  Altman, and Demler}}]{AltmanPNAS}
\bibinfo{author}{\bibfnamefont{A.}~\bibnamefont{Polkovnikov}},
  \bibinfo{author}{\bibfnamefont{E.}~\bibnamefont{Altman}}, \bibnamefont{and}
  \bibinfo{author}{\bibfnamefont{E.}~\bibnamefont{Demler}},
  \bibinfo{journal}{Proceedings of the National Academy of Sciences of the
  United States of America} \textbf{\bibinfo{volume}{103}},
  \bibinfo{pages}{6125} (\bibinfo{year}{2006}), ISSN \bibinfo{issn}{0027-8424},
  \href{http://www.pubmedcentral.nih.gov/articlerender.fcgi?artid=1458842\&tool=pmcentrez\&rendertype=abstract}{URL}.

\bibitem[{\citenamefont{Gritsev et~al.}(2006)\citenamefont{Gritsev, Altman,
  Demler, and Polkovnikov}}]{AltmanNP}
\bibinfo{author}{\bibfnamefont{V.}~\bibnamefont{Gritsev}},
  \bibinfo{author}{\bibfnamefont{E.}~\bibnamefont{Altman}},
  \bibinfo{author}{\bibfnamefont{E.}~\bibnamefont{Demler}}, \bibnamefont{and}
  \bibinfo{author}{\bibfnamefont{A.}~\bibnamefont{Polkovnikov}},
  \bibinfo{journal}{Nature Physics} \textbf{\bibinfo{volume}{2}},
  \bibinfo{pages}{705} (\bibinfo{year}{2006}), ISSN \bibinfo{issn}{1745-2473},
  \href{http://www.nature.com/doifinder/10.1038/nphys410}{URL}.

\bibitem[{\citenamefont{Fontanesi et~al.}(2010)\citenamefont{Fontanesi,
  Wouters, and Savona}}]{Fontanesi2010}
\bibinfo{author}{\bibfnamefont{L.}~\bibnamefont{Fontanesi}},
  \bibinfo{author}{\bibfnamefont{M.}~\bibnamefont{Wouters}}, \bibnamefont{and}
  \bibinfo{author}{\bibfnamefont{V.}~\bibnamefont{Savona}},
  \bibinfo{journal}{Phys. Rev. A} \textbf{\bibinfo{volume}{81}},
  \bibinfo{pages}{053603} (\bibinfo{year}{2010}),
  \href{http://link.aps.org/doi/10.1103/PhysRevA.81.053603}{URL}.

\bibitem[{\citenamefont{Cazalilla}(2004)}]{Cazalilla2004}
\bibinfo{author}{\bibfnamefont{M.~A.} \bibnamefont{Cazalilla}},
  \bibinfo{journal}{Journal of Physics B: Atomic, Molecular and Optical
  Physics} \textbf{\bibinfo{volume}{37}}, \bibinfo{pages}{S1}
  (\bibinfo{year}{2004}),
  \href{http://stacks.iop.org/0953-4075/37/i=7/a=051}{URL}.

\bibitem[{\citenamefont{Fontanesi et~al.}(2009)\citenamefont{Fontanesi,
  Wouters, and Savona}}]{Fontanesi2009}
\bibinfo{author}{\bibfnamefont{L.}~\bibnamefont{Fontanesi}},
  \bibinfo{author}{\bibfnamefont{M.}~\bibnamefont{Wouters}}, \bibnamefont{and}
  \bibinfo{author}{\bibfnamefont{V.}~\bibnamefont{Savona}},
  \bibinfo{journal}{Phys. Rev. Lett.} \textbf{\bibinfo{volume}{103}},
  \bibinfo{pages}{030403} (\bibinfo{year}{2009}),
  \href{http://link.aps.org/doi/10.1103/PhysRevLett.103.030403}{URL}.

\bibitem[{\citenamefont{Huhtam\"aki et~al.}(2007)\citenamefont{Huhtam\"aki,
  M\"ott\"onen, Ankerhold, and Virtanen}}]{Huhtamaki}
\bibinfo{author}{\bibfnamefont{J.~A.~M.} \bibnamefont{Huhtam\"aki}},
  \bibinfo{author}{\bibfnamefont{M.}~\bibnamefont{M\"ott\"onen}},
  \bibinfo{author}{\bibfnamefont{J.}~\bibnamefont{Ankerhold}},
  \bibnamefont{and} \bibinfo{author}{\bibfnamefont{S.~M.~M.}
  \bibnamefont{Virtanen}}, \bibinfo{journal}{Phys. Rev. A}
  \textbf{\bibinfo{volume}{76}}, \bibinfo{pages}{033605}
  (\bibinfo{year}{2007}).

\bibitem[{\citenamefont{Christiansen and Madsen}(2006)}]{Christiansen2006}
\bibinfo{author}{\bibfnamefont{S.}~\bibnamefont{Christiansen}}
  \bibnamefont{and} \bibinfo{author}{\bibfnamefont{P.~A.}
  \bibnamefont{Madsen}}, \bibinfo{journal}{Applied Numerical Mathematics}
  \textbf{\bibinfo{volume}{56}}, \bibinfo{pages}{91 } (\bibinfo{year}{2006}),
  ISSN \bibinfo{issn}{0168-9274},
  \href{http://www.sciencedirect.com/science/article/B6TYD-4FPX2G6-2/2/b73ab55719b73f208bb6b00b2b1d2a79}{URL}.

\end{thebibliography}

\end{document}